\documentclass[12pt]{article}
\usepackage{amsmath,amssymb,amsthm,latexsym,euscript,amscd, authblk}
\usepackage[english]{babel}
\usepackage[all]{xy}
\usepackage[dvips]{graphics}

\title{On the structutre of the algebra generated by the non-commutative operator graph demonstrating superactivation for a zero-error capacity}

\author[1]{G.G. Amosov\thanks {gramos@mi.ras.ru}}

\author[2, 3]{I.Yu. Zhdanovskiy\thanks {ijdanov@mail.ru}}

\affil[1]{Steklov Mathematical Institute of Russian Academy of Sciences}
\affil[2]{Moscow Institute of Physics and Technology}
\affil[3]{National Research University High School of Economics, Laboratory of Algebraic Geometry}

\begin{document}

\maketitle

\begin{abstract}
Recently M.E. Shirokov \cite{maxim} introduced the non-commutative operator graph depending on the complex parameter $\theta $ to construct channels with positive quantum zero-error capacity having vanishing n-shot capacity.
We study the algebraic structure of this graph. The relations for the algebra generated by the graph are derived. In the limiting case $\theta =\pm1$ the graph becomes abelian and degenerates into the direct sum of four one-dimensional irreducible representations of the Klein group.
\end{abstract}

\section{Introduction}

The superactivation of a quantum channel capacity was discovered in \cite{super}. It appeared that the quantum capacity for a tensor product of two quantum channels may be positive while the quantum capacity of each of these channels is zero. In \cite{graph1, graph2} it was shown that a value of the quantum capacity is closely related to the so-called non-commutative graph of
a quantum channel. The same phenomenon for the zero-error classical capacities was established in \cite {winter}. In \cite {shulman1, shulman2} some techniques for studying superactivation by
means of non-commutative graphs was introduced. It allows to give low-dimensional examples of superactivation for a quantum capacity. In our paper we shall analyse the algebra generated by the non-commutative graph introduced in \cite{maxim}.

\section{Preliminaries}

Denote $B(H)$ the algebra of all bounded operators and $\mathfrak {S}(H)$ the convex set of quantum states (positive unit trace operators) in a finite-dimensional Hilbert state $H$. Let $\Phi: B(H_A)\to B(H_B)$ be a quantum channel, i.e. a completely positive trace-preserving linear map transmitting $\mathfrak {S}(H_A)$ to the subset of $\mathfrak {S}(H_B)$ in the Hilbert spaces $H_A$ and $H_B$ \cite {holevo}.
Then the dual map $\Phi ^* : B(H_B)\to B(H_A)$ is defined by the formula
$$
Tr(\rho \Phi ^*(x))=Tr(\Phi (\rho)x),\ \rho \in \mathfrak {S}(H_A),\ x\in B(H_B).
$$
The Stinespring
theorem results in the representation of a channel $\Phi $ in the form
\begin{equation}\label{stine}
\Phi (\rho )=Tr_{H_E}V\rho V^*,\ \rho \in \mathfrak {S}(H_A),
\end{equation}
where $H_E$ is a Hilbert space of the environment and $V:H_A\to H_B\otimes H_E$ is an isometry. The representation (\ref {stine}) allows to define a complementary quantum
channel $\hat \Phi :\mathfrak {S}(H_A)\to \mathfrak {S}(H_E)$ by the formula \cite{holevo, compl}
\begin{equation}\label{compl}
\hat \Phi (\rho )=Tr_{H_B}V\rho V^*,\ \rho \in \mathfrak {S}(H_A).
\end{equation}
Taking into account (\ref {stine}) it is possible to derive the Kraus representation
\begin{equation}\label{kraus}
\Phi (\rho )=\sum \limits _{k=1}^{dimH_E}V_k\rho V_k^*,\ \rho \in \mathfrak {S}(H_A),
\end{equation}
where $\{V_k\}$ is a set of linear operators $V_k: H_A\to H_B$, $\sum \limits _jV_k^*V_k=I_{H_A}$. Using (\ref {kraus}) the complementary channel (\ref {compl}) can be represented as follows
\begin{equation}\label{compl2}
\hat \Phi (\rho )=\sum \limits _{j,k=1}^{dimH_E}Tr [ V_j\rho V_k ] |j><k|.
\end{equation}

The non-commutative graph \cite {graph1, winter} $\mathcal {G}(\Phi )$ of a quantum channel $\Phi $ is a closed linear span of the Kraus operators $\{V_j^*V_k\}_{j,k=1}^{dimH_E}$.
It follows from (\ref {compl2}) that $\mathcal {G}(\phi )$ is equal to the subspace $\hat \Phi ^*(B(H_E))$. The operator space $S$ is a non-commutative graph for some quantum
channel iff $x\in S$ implies $x^*\in S$ and the identity operator $I\in S$ \cite {graph1, shulman1}. In \cite {maxim} the following operator graph was introduced
\begin{equation}\label{L}
\begin{pmatrix}
a & b & c\theta & d\\
b & a & d & c/\theta\\
c/\theta & d & a & b\\
d & c\theta & b & a
\end{pmatrix}
\end{equation}
for $\theta \in {\mathbb C},\ |\theta |=1$. The operator graph (\ref {L}) is associated with pseudo-diagonal channels \cite {ruskai}
\begin{equation}\label{pseudo}
\Phi (\rho )=\sum \limits _{j,k}c_{j,k}<\psi _j|\rho |\psi _k>|j><k|
\end{equation}
where $\{c_{ij}\}$ is a Gram matrix of a collection of unit vectors, $\{|\psi_i>\}$ is a collection of vectors in $H$ such that $\sum \limits _i|\psi _i><\psi _i|=I$. Here we recover
the structure of the algebra ${\mathcal M}_{\theta }$ generated by the graph (\ref {L}) for any complex parameter $\theta \in \mathbb {C},\ \theta \neq 0$.

\section{The structure of the algebra ${\mathcal M}_{\theta }$ generated by the graph.}

It is straightforward to check that the graph (\ref {L}) is a linear envelope of the identity $I$ and the following three matrices:
\begin{equation}
\label{xyz}
X =
\begin{pmatrix}
0 & 1 & 0 & 0\\
1 & 0 & 0 & 0\\
0 & 0 & 0 & 1\\
0 & 0 & 1 &0
\end{pmatrix},
Y =
\begin{pmatrix}
0 & 0 & \theta & 0\\
0 & 0 & 0 & 1/\theta\\
1/\theta & 0 & 0 & 0\\
0 & \theta & 0 & 0
\end{pmatrix},
Z =
\begin{pmatrix}
0 & 0 & 0 & 1\\
0 & 0 & 1 & 0\\
0 & 1 & 0 & 0\\
1 & 0 & 0 & 0
\end{pmatrix}
\end{equation}
The matrices (\ref {xyz}) satisfy the relations
\begin{equation}\label{relations}
X^2=Y^2=Z^2=I,\ XZ=ZX,\ YZ=ZY.
\end{equation}
Let us consider the group $G$ generated by formal variables $x,y,z$ satisfying the relations: $x^2 = y^2 = z^2 = 1, xz=zx,yz=zy$. 
Note that adding to the last relations $xy=yx=z$ we obtain the Klein group.
Consider the subalgebra ${\cal M}_{\theta} \subset Mat_4({\mathbb C})$ generated by the matrices $X,Y,Z$. Thus, we have a well-defined morphism of algebras: $\phi_{\theta}: {\mathbb C}G \to {\cal M}_{\theta}$ defined by the rule: $x \mapsto X, y \mapsto Y, z \mapsto Z$. Also, the morphism $\phi_{\theta}$ defines a 4-dimensional representation of the group $G$.

{\bf Theorem 1.} {\it
We have the following statements:
\begin{itemize}
\item{If $\theta \ne \pm 1$ then the algebra ${\cal M}_{\theta}$ is a direct sum of two matrix algebras of size 2. In this case ${\rm dim}_{\mathbb C}{\cal M}_{\theta} = 8$.}
\item{If $\theta = \pm 1$ then the algebra ${\cal M}_{\theta}$ is isomorphic to the group algebra of the Klein group. In this case ${\rm dim}_{\mathbb C}{\cal M}_{\theta} = 4$.}
\end{itemize}
}

Proof.

Consider a normal subgroup $P \lhd G$ of index 2 generated by the elements $g = xy$ and $z$. Note that we have the following relation: $xgx = g^{-1}$. Using the relations of the group $G$, we obtain that the group $P$ is abelian. Hence, all irreducible representations of $P$ are 1-dimensional. Any 1-dimensional representation  is said to be a character of $P$. Thus, a dimension of irreducible representations of $G$ is less or equal to 2.
Let us describe characters of $P$. A character $\chi$ of $P$ is a morphism: $\chi :P \to {\mathbb C} \setminus \{0\},\ \chi (ab)=\chi (a)\chi (b),\ a,b\in P$. The character $\chi$ is determined by two numbers $\chi (g)\in {\mathbb C}\setminus \{0\}$ and $\chi (z),\ \chi (z)^2=1$. The last condition implies $\chi (z)=\pm 1$.

One can describe the standard construction of a $G$-representation $V_{\chi}$ induced by the character $\chi$ of $P$ as follows. Consider a vector $v$ such that $a v = \chi(a) v, a \in P$. Also, we can consider a "formal" vector $x \cdot v$. Thus, the vector space generated by $v$ and $x \cdot v$ is a space of $V_{\chi}$. Elements of $P$ act on $V_{\chi}$ by the rule: $a v = \chi(a) v$. Using the relations $x^2 = 1$ and $xax = a^{-1}$, we get $a (x\cdot v) = x (xa x \cdot v) = x a^{-1} \cdot v = \chi(a^{-1}) x \cdot v = \chi^{-1}(a) x \cdot v$. It means that $V_{\chi}$ as representation of $P$ is a direct sum of 1-dimensional representations corresponding to characters $\chi$ and $\chi^{-1}$.
The element $x$ acts by the formula: $x v = x \cdot v$ and $x x\cdot v = x^2 v = v$. Also, one can show that any 2-dimensional irreducible representation of $G$ is induced by a character of $P$.  
Consider the representation $\phi_{\theta}$ of the group algebra ${\mathbb C}G$ for $\theta \ne \pm 1$. In this case there are two irreducible submodules $V_{\chi}$ and $V_{-\chi}$ of representation $\phi_{\theta}$, where characters $\chi$ and $-\chi$ are defined by rule: $\chi(g) = \theta$ and $-\chi(g) = -\theta$. Using the standard arguments, we get that the $G$-representation $\phi_{\theta}$ is a direct sum $V_{\chi} \oplus V_{-\chi}$.  
It means that if $\theta \ne \pm 1$ then algebra ${\cal M}_{\theta}$ is a direct sum of the matrix algebras of size 2. Thus, we get the first statement. In the case $\theta = \pm 1$, one can check that the representation $\phi$ is a sum of four 1-dimensional representations.

$\Box $

We can see that there is a gap in dimension of the algebras ${\cal M}_{\theta}$ in the case $\theta = \pm 1$. We can change this situation as follows. We will construct a family of algebras ${\cal A}_{\theta}, \theta \in {\mathbb C} \setminus \{0\}$ and a morphism: $\psi: {\cal A}_{\theta} \to {\cal M}_{\theta}$ such that ${\rm dim}_{\mathbb C}{\cal A}_{\theta} = 8$ and $\psi: {\cal A}_{\theta} \cong {\cal M}_{\theta}$ for $\theta \ne \pm 1$. In the case $\theta = \pm 1$ morphism $\psi:{\cal A}_{\theta} \to {\cal M}_{\theta}$ is a surjective and ${\rm Ker}\psi$ is a 4-dimensional nilpotent ideal of ${\cal A}_{\theta}$. Roughly speaking, we construct family of algebras ${\cal A}_{\theta}$ isomorphic to ${\cal M}_{\theta}$ for general $\theta$. In the case $\theta = \pm 1$ algebras ${\cal A}_{\theta}$ "degenerates" to ${\cal M}_{\theta}$.

Consider the algebra ${\cal A}_{\theta}$ generated by "formal" variables $x,y,z$ and the relations: $x^2 = y^2 = z^2 = 1, xz = zx, yz = zy, xy + yx = (\theta + \theta^{-1})z$. We can rewrite the last relation in the following manner: $g + g^{-1} = (\theta + \theta^{-1})z$, where $g=xy$.
One can check that this relation is true for the matrices $X,Y,Z$. Thus, we have a surjective morphism: $\psi: {\cal A}_{\theta} \to {\cal M}_{\theta}$ defined by the rule: $\psi:x \mapsto X, y \mapsto Y, z \mapsto Z$.
We can prove the following theorem:

{\bf Theorem 2.} {\it
\begin{itemize}
\item{If $\theta \ne \pm 1$ then $\psi$ is isomorphism.}
\item{If $\theta = \pm 1$ then the algebra ${\cal A}_{\theta}$ has the 4-dimensional two-sided nilpotent ideal $J$ such that $\psi: {\cal A}_{\theta}/J \cong {\cal M}_{\theta}$.}
\end{itemize}
}

Proof.

Let us show that a dimension of the algebra ${\cal A}_{\theta}$ is equal to 8. Actually, one can show that if $\theta \ne \pm i$ the algebra ${\cal A}_{\theta}$ has the following basis: $1, g, g^2, g^3, x, xg, xg^2, xg^3$, where $g = xy$. In the case $\theta = \pm i$ one can show that the algebra ${\cal A}_{\theta}$ has the basis: $1,g,x,z,xg,xz,gz,xgz$. Thus, in the case $\theta \ne \pm 1$, $\psi$ is bijective, hence, $\psi$ is isomorphism.

Consider the case $\theta = \pm 1$. In this case we have the relation: $g + g^{-1} = \pm 2 z$. Thus, $(g+g^{-1})^2 = 4z^2 = 4$. And, hence, we obtain the following relation:
\begin{equation}
\label{nilp}
(g^2 - 1)^2 = 0.
\end{equation}
Consider the ideal $J$ of the algebra $A$ generated by $g^2 - 1$. One can see that $J$ has the following basis $g^2 - 1, x(g^2 - 1), g(g^2 - 1), xg(g^2 - 1)$. Also, consider $J^2 = \langle t_1 t_2, t_i \in J \rangle$. Using the relation (\ref{nilp}), we get $J^2 = 0$. The ideal satisfying this property is said to be a radical. Therefore, the algebra ${\cal A}_{\theta}$ for $\theta = \pm 1$ has a 4-dimensional radical. One can check that $\psi(J) = 0$.

$\Box $

\section{Conclusion} We have found the general features of the algebra generated by the non-commutative operator graph playing an important role in quantum information theory. We have presented only a sketch of the theory.
We are planning to continue the study in the future to discover the algebraic nature of the quantum superactivation.

\section*{Acknowledgments} The authors are grateful to A.S. Holevo for kind attention to the work and many useful remarks. Theorem 1 was proved by G.G. Amosov, Theorem 2 was proved by
I.Yu. Zdanovsky. The work of G.G. Amosov is
supported by Russian Science Foundation under grant No 14-21-00162 and performed in Steklov Mathematical Institute of Russian Academy of Sciences.
The work of I.Yu. Zhanovskiy is supported by RFBR, research projects 13-01-00234 and 14-01-00416, and was prepared within the framework of
a subsidy granted to the HSE by the Government of the Russian Federation for the implementation of the Global Competitiveness Program.

\end{document}